\documentclass[]{JHEP3}
\usepackage{epsfig}
\usepackage{graphicx}
\def\be{\begin{equation}}
\def\ee{\end{equation}}
\def\bea{\begin{array}}
\def\eea{\end{array}}
\def\beqa{\begin{eqnarray}}
\def\eeqa{\end{eqnarray}}
\def\beqas{\begin{eqnarray*}}
\def\eeqas{\end{eqnarray*}}

\def\bp{\begin{picture}}
\def\ep{\end{picture}}
\def\bc{\begin{center}}
\def\ec{\end{center}}
\def\bfig{\begin{figure}}
\def\efig{\end{figure}}

\def\bit{\begin{itemize}}
\def\eit{\end{itemize}}
\def\nn{\nonumber}
\def\f{\frac}

\def\[{\left[}
\def\]{\right]}
\def\({\left(}
\def\){\right)}

\def\..{\left.}
\def\.{\right.}
\def\tl{\tilde}
\def\ra{\rightarrow}
\def\la{\leftarrow}

\def\tm{\times}

\def\da{\dagger}

\def\la{\lambda}

\def\al{\alpha}

\def\ep{\epsilon}

\def\ga{\gamma}
\def\pa{\partial}
\def\pr{\prime}

\title{ Interpreting The 750 GeV Diphoton Excess Within Topflavor Seesaw Model}
\author{Junjie Cao$^{1}$, Liangliang Shang$^{1}$, Wei Su$^{2}$, Fei Wang$^{3,2}$, Yang Zhang$^{3}$\\
$^1$  Department of Physics, Henan Normal University, Xinxiang 453007, P.R. China\\
$^2$ State Key Laboratory of Theoretical Physics, Institute of Theoretical Physics, Chinese Academy of Sciences, Beijing 100080, P. R. China
$^3$ School of Physics, Zhengzhou University, ZhengZhou 450000, P.R.China\\
}
\abstract{We propose to interpret the 750 GeV diphoton excess in a typical topflavor seesaw model. The new resonance $X$ can be
identified as a CP-even scalar emerging from a certain bi-doublet Higgs field. Such a scalar can couple to charged scalars, fermions as
well as heavy gauge bosons predicted by the model, and consequently all of the particles contribute to the diphoton decay mode of the X.
Numerical analysis indicates that the model can predict the central value of the diphoton excess without contradicting
any constraints from 8 TeV LHC, and among the constraints,
the tightest one comes from the $Z \gamma$ channel, $\sigma_{8 {\rm TeV}}^{Z \gamma} \lesssim 3.6 {\rm fb}$, which requires
$\sigma_{13 {\rm TeV}}^{\gamma \gamma} \lesssim 6 {\rm fb}$ in most of the favored parameter space.}

\begin{document}
\maketitle \indent
\newpage
\section{Introduction}
Recently in the searches for new physics at the LHC Run-II with $\sqrt{s}=13 {\rm TeV}$ and  3 fb$^{-1}$ integrated data,
 both the ATLAS and CMS collaborations reported a diphoton excess with an invariant
mass around 750 GeV \cite{ATLAS:750,CMS:750}. Combined with the 8 TeV data, the favored rate of the excess is given by
\begin{equation}
\sigma^{750 GeV}_{\gamma\gamma} = (4.4\pm 1.1) ~\rm{fb}~.\label{excess}
\end{equation}
in the narrow width approximation \cite{ex-0}.
Although the local significance is not very high, which is only $3.9\sigma$ for ATLAS data and $2.6 \sigma$ for CMS data, this excess
was widely regarded as a possible hint of new physics beyond the Standard Model (SM).

So far about one hundred theoretical papers have appeared to interpret the excess in various models
\cite{ex-0,ex-00,ex-1,ex-2,ex-3,ex-4,ex-5,ex-6,ex-7,ex-8,ex-9}, and most of them employed
the process $g g \to X \to \gamma \gamma$ with $X$ denoting a  $750 \ {\rm GeV}$ scalar particle to fit the data. From
these studies, one can infer two essential properties of the $X$. One is that its interactions with the SM particles
other than gluons and photons should be significantly weaker than those of the SM Higgs boson.
In this case, the rates of the $X$-mediated processes $p p \to X \to ZZ, W W^\ast, h h, f\bar{f}$ are suppressed so that no excess
on these channels was observed at the LHC Run-I \cite{ex-9}. The other is that the $X$ must interact with new charged and colored
particles to induce the effective $X\gamma \gamma$ and $X gg$ couplings through their loop effects. In order to explain the excess
in a good way, the new particles should be lighter than about $1\  {\rm TeV}$, and meanwhile their interactions with
the $X$ must be rather strong.

Among the new physics models employed to interpret the excess, the minimal theoretical framework is the extension of the SM by one gauge singlet scalar
field and vector-like fermions \cite{ex-00}. This model was extensively discussed since it provides a very simple but meanwhile feasible explanation of the excess.
However, as pointed out in \cite{vacuum}, in order to explain the excess the Yukawa couplings of the fermions are usually so large
that the vacuum state of the scalar potential becomes unstable at a certain energy scale, which implies that other new physics must exist.
This motivates us to speculate what is the fundamental theory behind the minimal model.  We note that for the interaction $\lambda_F X \bar{F} F$ with $F$ denoting
a vector-like fermion, its contribution to the $X\gamma \gamma$ coupling is determined by the ratio $\frac{\lambda_F}{M_F}$ under
the condition $4 M_F^2 \gg (750 {\rm GeV})^2$. In a usual theory, $\frac{\lambda_F}{M_F} \sim \frac{1}{v}$ with $v \sim 1\ {\rm TeV}$
denoting the typical size for both the fermion mass and other new particle masses in the theory. On the other hand, if the fermion acquires its mass in a complicated way
and consequently $\lambda_F$ and $M_F$ are less correlated, the ratio can be much larger than $1/v$ with $v \gg M_F$ denoting the mass scale of
the other new particles.  For such a situation, the effective theory at the TeV scale contains only the scalar $X$ and the fermions,
which is similar to the minimal model, but it usually has a more complicated scalar sector than the minimal model.
Building such a model and using it to explain the diphoton excess are the aims of this work. Obviously, such a study can improve
our understanding on the minimal framework.

To actualize the idea, we are motivated by the top-specific theories, such as the top condensation models \cite{tcond,topcolor},
the top seesaw model \cite{topseesaw,HHT} and the topflavor model \cite{topflavor}, which attempts to
interpret the relatively large top quark mass in comparison with the other SM
particles.  Roughly speaking, we assume that the third generation fermions undergo a different
$SU(2)$ weak interaction from the first two generation fermions \cite{flavor-dynamics}. At the same time, we introduce new vector-like fermions and
split their mass spectrum by seesaw mechanism \cite{hjhe:topflavorseesaw}.  In this way, the $750 {\rm GeV}$ resonance
is identified as a CP-even scalar emerging from a bi-doublet Higgs, which triggers the breaking of the two SU(2) gauge symmetry into $SU(2)_L$, and its
interactions with photons are induced by relevant scalars, fermions as well as gauge bosons.  Due to these features,
our model has more freedom to explain the diphoton excess than the minimal model. We remind that our model is somewhat similar to
the topflavor seesaw model proposed in \cite{hjhe:topflavorseesaw}, so we dub it hereafter the typical topflavor seesaw model.

This paper is organized as follows. In section 2, we introduce the structure of the typical top flavor seesaw model, and list its
particle spectrum. In section 3, we choose benchmark scenarios to study the diphoton excess. Subsequently we draw
conclusions in section 4. We also present more details of our model in the Appendix.

\section{The framework of the topflavor seesaw model}

In this section, we recapitulate the structure of the typical topflavor seesaw model. This model is based on the gauge symmetry group
$SU(3)_c\tm SU(2)_1\tm SU(2)_2 \tm U(1)_Y$, where the third generation of fermions transform non-trivially under the
$SU(2)_2$ group, while the first two generation of fermions transform by the $SU(2)_1$ group.
The symmetry breaking of the gauge group into the electromagnetic group $U(1)_{Q}$ is a
two-stage mechanism: first, $SU(2)_1 \times SU(2)_2 \times U(1)_Y$ breaks down into $SU(2)_L \times U(1)_Y$
at the TeV scale; second, $SU(2)_L \times U(1)_Y$ breaks down into the $U(1)_{Q}$ at the electorweak scale.
This breakdown can be accomplished by introducing two Higgs doublets and one bi-doublet Higgs
with the following $SU(3)_c\tm SU(2)_1\tm SU(2)_2 \tm U(1)_Y$ quantum numbers
\beqa
H_1\sim(~1,~2,~1)_{-1/2}~,~H_2\sim(~1,~1,~2)_{-1/2}~,~\Phi\sim(~1,~2,~2)_{0}.
\eeqa
In this model, we also introduce following vector-like quarks and leptons to couple with the bi-doublet
\beqa
&&V_L\equiv({\cal{T}},{\cal{B}})_L\sim (3,2,1)_{1/6}~,~~V_R\equiv({\cal{T}},{\cal{B}})_R\sim (3,1,2)_{1/6}~,\nn\\
&&{V}^\pr_R\equiv(\tl{\cal{T}},\tl{\cal{B}})_R\sim (3,{2},1)_{1/6}~,~~{V}^\pr_L\equiv(\tl{\cal{T}},\tl{\cal{B}})_L\sim (3,1,{2})_{1/6}~,\nn\\
&&\tl{V}_L\equiv({\cal{N}},{\cal{E}})_L\sim (1,2,1)_{-1}~,~~\tl{V}_R\equiv({\cal{N}},{\cal{E}})_R\sim (1,1,2)_{-1}~,\nn\\
&&\tl{V}^\pr_R\equiv(\tl{\cal{N}},\tl{\cal{E}})_R\sim (1,{2},1)_{-1}~,~~\tl{V}^\pr_L\equiv(\tl{\cal{N}},\tl{\cal{E}})_L\sim (1,1,{2})_{-1}~,
\eeqa
and write their mass terms as follows
 \beqa
 -{\cal L}\supseteq  M_H \bar{V}_L V^\pr_R + M_H \bar{V}_R V^\pr_L+\bar{V}_L\(\la_V\Phi\) V_R+\bar{V}^\pr_L\(\la_V\Phi\) V^\pr_R+ ({\rm V,V^\pr\ra \tl{V},\tl{V}^\pr}),
 \eeqa
where the dimensionful coefficient $M_H$ may have a dynamical origin or just be imposed by hand.

With the above field settings, the remaining Lagrangian is
 \beqa
{\cal L} \supseteq && {\cal L}_{kin} + {\cal L}_Q+{\cal L}_L - V(\Phi)-V(H_1,H_2) ~, \label{Fermion-mass} \\
-{\cal L}_Q \supseteq && \bar{Q}_{3,L} \(H_2 y_tt_R+i\tau_2 H_2^*y_b b_R\)+  \bar{V}_{L} \(H_1 y_{T} t_R+i\tau_2H_1^*y_{B} b_R\)+m_{V,Q} \bar{Q}_{3,L} V_R  \nn\\
&&+ \bar{V}_{L}^\prime \(H_2 y_{T}^\prime t_R+ i \tau_2H_2^*y_{B}^\prime b_R\) + \sum\limits_{i=1}^2 \bar{Q}_{i,L} \(H_1 y_u^i u^R_i +i\tau_2 H_1^*y_d^i d_i^R\) + h.c.~,\nn\\
-{\cal L}_L \supseteq & & \bar{L}_{3,L} \(H_2 y_{\nu_\tau} \nu_{\tau,R}+i\tau_2 H_2^*y_\tau \tau_R\)+\bar{\ \tl{V}_{L}}\(H_1y_N \nu_{\tau,R}+ i\tau_2 H_1^*y_E \tau_R\)+
m_{V,L} \bar{L}_{3,L} \tl{V}_R \nn \\
&&+\bar{\ \tl{V}_{L}^{\prime}} \(H_2 y_N^\prime \nu_{\tau, R} + i\tau_2 H_2^*y_E^\prime \tau_R\) + \sum\limits_{i=1}^2 \bar{L}_{i,L} \(H_1 y_N^i N_R^i +i\tau_2H_1^*y_E^i E_i^R\) + h.c.~, \nn
\eeqa
where $V(\Phi)$ represents the potential of the field $\Phi$, $V(H_1,H_2)$ corresponds to the potential of a general two Higgs doublet model,
$y_\alpha$ with $\alpha =t, b, \cdots$ and $y_\beta^\prime$ with $\beta = T, B, N, E$ are Yukawa coupling coefficients, and $m_{V,Q}$ and $m_{V,L}$ are dimensionful
parameters. The general expression of $V(\Phi)$ is given by \cite{Bi-doublet}
\beqa
V(\Phi)&=&-\mu_1^2 Tr(\Phi^\da\Phi)-\mu_2^2\[Tr(\tl{\Phi}\Phi^\da)+Tr(\tl{\Phi}^\da\Phi)\]+\la_1\[Tr(\Phi^\da\Phi)\]^2\nn\\
&&+\la_2\[Tr(\Phi^\da\tl{\Phi})Tr(\tl{\Phi}^\da{\Phi})\]+\la_3\(\[Tr(\Phi^\da\tl{\Phi})\]^2+\[Tr(\tl{\Phi}^\da{\Phi})\]^2\)\nn\\
&&+\la_4\(Tr(\Phi^\da\Phi)\[Tr(\tl{\Phi}\Phi^\da)+Tr(\tl{\Phi}^\da\Phi)\]\), \label{Bi-doublet-potential}
\eeqa
with $\tl{\Phi}=\sigma_2\Phi^*\sigma_2$.

About the Lagrangian in Eq.(\ref{Fermion-mass}), two points should be noted. One is that the field $\Phi$ may have a different dynamical origin from
that of the fields $H_1$ and $H_2$. So in  writing down the scalar potential, we neglect the couplings between $\Phi$ and $H_1,H_2$.
The other is that, the third generation fermions, which are specially treated in our model, can in principle mix with the vector-like quarks. If any of $y_\beta$, $y^\prime_\beta$ and $m_V$ is large, the flavor mixing of the SM fermions may differ significantly from its
SM prediction \cite{hjhe:topflavorseesaw}. Although this situation may still be allowed by the precise measurements in flavor physics, we require
all the coefficients to be sufficiently small to suppress the decay of the $750 {\rm GeV}$ resonance into the state $t \bar{t}$. We will turn to this issue later.

In the following, we list the spectrum of the particles that we are interested in.

\subsection{Scalar sector}

In the topflavor seesaw model, the bi-doublet Higgs contains 8 real freedoms, and it can parameterized by
\beqa
\Phi=\f{1}{\sqrt{2}}\(\bea{cc}\sqrt{2}v+\(\rho_1+i\eta_1\)&\sqrt{2} V_1^+\\ \sqrt{2} V_2^-&\sqrt{2}v+\rho_2+i\eta_2\eea\),
\eeqa
where $\rho_1$ and $\rho_2$ are CP-even fields with $\sqrt{2} v$ being their vacuum expectation value (vev),
$\eta_1$ and $\eta_2$ are CP-odd fields and $V_1^+$ and $V_2^-$ denote charged fields. The non-zero $v$ triggers
the breaking of the group $SU(2)_1\tm SU(2)_2$ into the diagonal $SU(2)_L$ group. In such a process,
the field combinations $\eta_1-\eta_2$ and $V_1^+-(V_2^-)^*$  act as Goldstone modes, and are absorbed
by the gauge bosons corresponding to the broken $SU(2)_H$ group.
Their orthogonal combinations correspond to the charged and CP-odd scalars respectively, which are given by
\beqa
H^+&=&\f{1}{\sqrt{2}}\[V_1^++(V_2^-)^*\]~, \\
A^0&=&\f{1}{\sqrt{2}}\[\eta_1+\eta_2\].
\eeqa
As for the CP-even fields $\rho_1$ and $\rho_2$, they mix to form mass eigenstates $h_0$ and $H_0$ in following way
\beqa
 \(\bea{c}h^0\\ H^0\eea\)=\frac{1}{\sqrt{2}} \(\bea{cc}~~1~~ &~~ 1~~ \\~ -1~ & ~~1~~\eea\) \(\bea{c} \rho_1\\ \rho_2 \eea\).
\eeqa
With the physical states, the field $\Phi$ can be reexpressed as
\beqa
\Phi \sim \f{1}{2}\(\bea{cc} 2 v+ h^0 - H^0 + i A^0& H^+\\ H^-& 2 v + h^0 + H^0 + i A^0\eea\).
\eeqa
This form is useful to understand our expansion result of the $V(\Phi)$.

From Eq.(\ref{Bi-doublet-potential}), one can get the minimization condition of the potential, $4 \kappa v^2=\mu_1^2+2\mu_2^2$ with
$\kappa = \la_1+ \la_2 + 2\la_3 + 2 \la_4 $, and the vacuum stability condition $\kappa >0$. One can
also get the spectrum of the scalars as follows
\begin{eqnarray}
m_{h^0}^2 & = & \mu_1^2 + 2 \mu_2^2 = 4 \kappa v^2,  \\
m_{H^0}^2 &=& 2 \mu^2_2 -  \mu^2_1 + 4 (\lambda_1 - \lambda_2 - 2 \lambda_3) v^2, \\
m_{A^0}^2 &=& 2 \mu^2_2 -  \mu^2_1 + 4 (\lambda_1 + \lambda_2 - 6 \lambda_3) v^2,  \\
m_{H^+}^2 & = & m_{H^0}^2 \label{m-H+}.
\end{eqnarray}

In a similar way, the two Higgs doublets $H_1$ and $H_2$ can be written as
\beqa
\langle H_i \rangle=\(H_i^+, \frac{1}{\sqrt{2}} (v_i + H_i^0 + i A_i^0) \)
\eeqa
with $v_1^2+v_2^2=v_{EW}^2 $ and $\tan\beta \equiv v_2/v_1$, and the non-zero $v_i$s break the $SU(2)_L \tm U(1)_Y$ gauge
symmetry into the $U(1)_Q$ symmetry. In this process, the alignment of the fields $H_1^0$ and $H_2^0$ forms a lightest
CP-even scalar, which corresponds to the 125 GeV higgs boson discovered by the ATLAS and CMS
collaborations at the LHC \cite{ATLAS:125,CMS:125}.

Throughout this work, we identify the state $h^0$ from $\Phi$ as the 750 GeV resonance, which is responsible
for the diphoton excess by the parton process $ g g \to h^0 \to \gamma \gamma$. Since we have neglected the mixing
between $H_i$ and $\Phi$, we do not consider the decay of $h^0$ into the SM-like Higgs boson pair. In fact, if we switch
on the mixing, the upper bound on the di-Higgs signal at the LHC Run I has required the mixing to be small \cite{ex-9}.

\subsection{Gauge bosons}

In our theory, the covariant derivative that appears in the kinetic term of $\Phi$ is given by
\begin{eqnarray}
D_\mu \equiv \partial_\mu - i h_1 W_{1,\mu}^a (T^a_1) + i h_2 W_{2,\mu}^b (T^b_2) + i g_Y Y B_\mu~,
\end{eqnarray}
where $T_1^a$ and $T_2^b$ with $a,b = 1,2,3$ are the $SU(2)$ generators, $Y$ is the hypercharge generator,
and $h_1$, $h_2$ and $g_Y$ are gauge coupling coefficients. After the first step symmetry breaking,
the $SU(2)_L$ coupling coefficient $g_2$ is related with $h_1$ and $h_2$ by
\beqa
\f{1}{g_2^2}=\f{1}{h_1^2}+\f{1}{h_2^2}~,  \label{gauge-coupling}
\eeqa
which implies $h_1 = g_2/\cos \theta$, $h_2 = g_2/\sin \theta$ with $\tan \theta \equiv h_1/h_2$,
and the gauge fields corresponding to the broken $SU(2)_H$ group (usually called flavoron and denoted by
$F_\mu^i$ hereafter) and the $SU(2)_L$ group (denoted by $W_\mu^i$) are
\beqa
 \(\bea{c} F^i_{\mu}\\ W^i_{\mu} \eea\)=\(\bea{cc}\sin\theta&\cos\theta\\-\cos\theta&\sin\theta\eea\) \(\bea{c}W^i_{1,\mu}\\ W^i _{2,\mu}\eea\)
 \eeqa
where $i=\pm, 3$. At this stage, the fields $F_{\mu}^\pm$ and $F_{\mu}^3$ are massive with a common squared mass of $(h_1^2+h_2^2)v^2= 4g_2^2 v^2 (\csc^2 2\theta)$,
and by contrast all the fields $W_{\mu}^i$ keep massless.

After the second step symmetry breaking, the masses of the fields $F_{\mu}^\pm $ keep unchanged, but the field $F^3_{\mu}$ mixes with the other neutral
gauge fields to form mass eigenstates. In the basis $(F_{\mu}^3,W_{\mu}^3,B_{\mu})$,
the squared mass matrix is given by
\beqa
\frac{h^2}{4} \(\bea{ccc} 4 v^2 + s_\theta^4 v^2_{1}+ c_\theta^4 v_2^2~~&~~ s_\theta c_\theta (s_\theta^2 v_1^2- c_\theta^2 v_2^2)~~&
~~-\f{g_Y}{4 h} (s_\theta^2 v_{1}^2 - c_\theta^2 v_2^2) \\ s_\theta c_\theta (s_\theta^2 v_1^2- c_\theta^2 v_2^2)~~&~~s_\theta^2 c_\theta^2 v_{EW}^2~~&~~-\f{g_Y}{h}
s_\theta c_\theta v_{EW}^2  \\-\f{g_Y}{4 h} (s_\theta^2 v_{1}^2 - c_\theta^2 v_2^2)~~&~~-\f{g_Y}{h}
s_\theta c_\theta v_{EW}^2 ~~&~~\f{g_Y^2}{h^2} v_{EW}^2 \eea\).
\eeqa
where $h = \sqrt{h_1^2 + h_2^2}$, $s_\theta \equiv \sin \theta$ and $c_\theta \equiv \cos \theta$.

This matrix can be diagonalized by a rotation $U$ to get mass eigenstates $(Z^\prime, Z, \gamma)$. Consequently we have
\beqa
\(Z^\pr~,Z~,\gamma\)= \(F_{\mu}^3~,W_{\mu}^3~,B_{\mu}\)U^T~.
\eeqa

\subsection{Heavy Fermions}

After the first step gauge symmetry breaking, the mass matrix of the vector-like fermions $V$ and $V^\pr$ are given by
\beqa
( \bar{V}_L~,\bar{V}^\pr_L)\(\bea{cc}\la_V v~ & ~M_H\\M_H~ &\la_V v\eea\)\(\bea{c}V_R\\V_R^\pr\eea\) + h.c..
\eeqa
with the eigenvalues $M_{1,2}=\la_V v\mp M_H$. The corresponding eigenstates are given by the combinations
\beqa
Q_{L,R}^1&=&\f{1}{\sqrt{2}}\(V_{L,R} - V_{L,R}^\pr\) \equiv ({\cal{T}}^1,{\cal{B}}^1)_{L,R}, \\
Q_{L,R}^2&=&\f{1}{\sqrt{2}}\(V_{L,R} + V_{L,R}^\pr\) \equiv({\cal{T}}^2,{\cal{B}}^2)_{L,R}.
\eeqa

In the basis $(t, {\cal{T}}^1, {\cal{T}}^2)$, the mass matrix of the heavy up-type quarks
at the weak scale is given by
\beqa
\(~\bar{t}_L~~\bar{{\cal{T}}}_L^1~~\bar{{\cal{T}}}_L^2 \)\(\bea{ccc} y_t v_2~~&~~\frac{1}{\sqrt{2}} m_V~~
&~~\frac{1}{\sqrt{2}} m_V  \\ \frac{1}{\sqrt{2}} (y_T v_1 - y_T^\prime v_2) ~~ &~~ \la_V v - M_H ~~& ~~0~~ \\
\frac{1}{\sqrt{2}} (y_T v_1 + y_T^\prime v_2) ~~ &~~0~~&~~ \la_V v + M_H \eea\) \left(\bea{c}t_R\\{\cal{T}}_R^1\\ {\cal{T}}_R^2 \eea\right)~.
\eeqa
This matrix can be diagonalized to get the mass eigenstates $(t_1, t_2, t_3)$. {Since we are interested in the case that
$ y_t v_2, m_V, (y_T v_1 + y_T^\prime v_2) \ll \la_V v - M_H < \la_V v + M_H $, the mass eigenstate $t_2$ is dominated by the field
${\cal{T}}^1$ with $m_{t_2} = \la_V v - M_H $, and  $t_3$ is dominated by the field
${\cal{T}}^2$ with $m_{t_3} = \la_V v + M_H $.}
We emphasize that the mixings between $t$ and ${\cal{T}}^i$ can induce the $h^0 \bar{t}_1 t_1$, $H^0 \bar{t}_1 t_1$ and $H^+ \bar{t}_1 b$
interactions with $t_1$ identified as the top quark measured in experiments. In our discussion about the diphoton excess, we assume
the mixings to be sufficiently small so that $Br(h^0 \to g g) \gg Br (h^0 \to t_1 \bar{t}_1)$,
and consequently we neglect the contribution of $h^0 \to t_1 \bar{t}_1$ to the total width of $h^0$. The same mixings can also induce the decay
$t_2, t_3 \to W b, t_1 Z$, and the LHC searches for
vector-like fermions have required $m_{t_{2,3}} \gtrsim 800 {\rm GeV}$ \cite{vector-like-quark-search}.

Note that similar discussions can be applied to the down-type quarks and leptons.

\section{The diphoton excess}

If the diphoton excess observed by both ATLAS and CMS collaborations is initiated by gluon fusion, its production rate can be written as \cite{ex-9,giudice}
\begin{eqnarray}
\sigma^{\gamma \gamma}_{13 TeV} (p p \to h^0 \to \gamma \gamma) = \frac{\Gamma_{h^0 \to gg}}{\Gamma^{SM}_{H \to g g}} |_{m_H \simeq 750 {\rm GeV}} \times \sigma^{SM}_{\sqrt{s}=13 {\rm TeV}} (H) \times Br (h^0 \to \gamma \gamma),
\label{cross-section}
\end{eqnarray}
where $\Gamma_{h^0 \to g g}$ is the width for the decay $h^0 \to g g$, $\Gamma^{SM}_{H \to g g} = 6.22 \times 10^{-2} {\rm GeV}$ denotes the width
of the SM Higgs $H$ decay into $g g$ with $m_{H} = 750 {\rm GeV}$ and $\sigma^{SM}_{\sqrt{s}=13 {\rm TeV}} (H)=735 \ {\rm fb}$  is the NNLO production rate of the $H$ at the 13 TeV LHC \cite{HCS}. As pointed out in \cite{ex-0}, after combining the diphoton data at the 13 TeV LHC with those at the 8 TeV LHC,
the preferred rate for the excess at the 13 TeV LHC is
\begin{equation}
\sigma^{\gamma \gamma}_{13 TeV} (p p \to h^0 \to \gamma \gamma ) = (4.6\pm 1.2) ~\rm{fb}. \label{excess}
\end{equation}
This rate can be transferred to the requirement
\begin{eqnarray}
\frac{\Gamma_{h^0 \to g g}}{\Gamma_{tot}} \times \Gamma_{h^0 \to \gamma \gamma} = (3.9 \pm 1.0 ) \times 10^{-4} {\rm GeV},  \label{requirment}
\end{eqnarray}
where $\Gamma_{tot}$ denotes the total width of $h^0$, and in the topflavor seesaw model it is given by
\begin{eqnarray}
\Gamma_{tot} = \Gamma_{h^0 \to g g} + \Gamma_{h^0 \to \gamma \gamma} + \Gamma_{h^0 \to Z \gamma} + \Gamma_{h^0 \to W W^\ast} +
\Gamma_{h^0 \to Z Z}.
\end{eqnarray}

\subsection{Useful formulae for calculation}

In this part, we list the formulae for the partial widths to calculate the $\Gamma_{tot}$.
\begin{itemize}
\item The widths of $h^0 \to \gamma \gamma, g g$ are given by
\begin{eqnarray}
\Gamma_{h^0 \ra \gamma\gamma} &=&\f{\al^2 m_{h^0}^3}{1024\pi^3}\left| I^{h^0}_{\gamma \gamma} \right |^2, \quad \quad
\Gamma_{h^0 \ra gg} = \f{\al_S^2 m_{h^0}^3}{32 \pi^3}\left| I^{h^0}_{gg} \right |^2,
\end{eqnarray}
where $I^{h^0}_{\gamma \gamma}$ and $I^{h^0}_{g g}$ parameterize the $h^0 \gamma \gamma $ and $h^0 g g$ interactions,
and their general expressions are \footnote{We remind that
the signs for the second and third terms in the expression of $\Gamma_{h^0 \ra \gamma\gamma}$ are opposite to those in \cite{carena}.
This is due to the sign convention, and it does not affect the results in this work. }
\begin{eqnarray}
I^{h^0}_{\gamma \gamma} &=& \f{g_{h^0VV}}{m_V^2} N_{c, V} Q_V^2A_1(\tau_V) - \f{2g_{h^0FF}}{m_F} N_{c,F}Q_F^2 A_{1/2}(\tau_F) -
\f{g_{h^0SS}}{m_S^2}N_{c,S}Q_S^2A_0(\tau_S), \nonumber \\
I^{h^0}_{g g} &=& \frac{1}{2} \f{g_{h^0FF}}{m_F} A_{1/2}\(\f{4 m_F^2}{m_S^2}\).
\end{eqnarray}
In above expressions, the coefficient $g_{h^0 X X}$ with $X= V, F, S$ represents the coupling of the $h^0 X^\ast X$ interaction,
$m_X$, $N_{c, X}$ and $Q_X$  are the mass, color number and electric charge of the particle $X$ respectively, and $\tau_X=4m_X^2/m_{h^0}^2$.
The involved loop functions are defined by \cite{carena}
 \beqa
 A_{1}(x)&=&-[2+3x+3(2x-x^2)f(x)]~,\nn\\
  A_{\f{1}{2}}(x)&=& 2x [1+(1-x)f(x)]~,\nn\\
  A_0(x)&=&-x(1-x f(x))~,\nn\\
 f(x)&=& arcsin^2\(\f{1}{\sqrt{x}}\),~~~~x\geq 1.
 \eeqa

Obviously, the three terms in $I^{h^0}_{\gamma\gamma}$ correspond to the contributions from vector bosons, fermions and scalars, respectively.
In our model, they are given by: gauge bosons $V = F_\mu^+,F_\mu^-$, fermions $F = t_2, t_3, b_2, b_3, \tau_2, \tau_3$ and scalars $S = H^+,H^-$.  In the appendix, we present all couplings used in our calculation, including the expressions of $g_{h^0 V V}$, $g_{h^0 F F}$ and $g_{h^0 S S}$.

\item The width for $h^0 \to Z \gamma$ can be obtained in a way quite similar to that for $h^0 \to \gamma \gamma$, and it is given by \cite{ex-9}
\begin{eqnarray}
\Gamma_{h^0\to Z\gamma}
=\frac{G^2_F m^2_W \alpha m_{h^0}^3}{64 \pi^4}
  \left( 1 - \frac{m^2_Z}{m^2_{h^0}} \right)^3
  \left|I_{Z \gamma}^{h^0}\right|^2,
\end{eqnarray}
where
\begin{eqnarray}
\nonumber
I_{Z\gamma}^{h^0} &=& \frac{m_W}{g^2_2}\left [ \frac{ g_{h^0VV} g^\prime_{ZVV} }{ m^2_V } N_{c,V} Q_V  A_1(\tau_V)
- \frac{ 2 g_{h^0FF} g^\prime_{ZFF} }{ m_F } N_{c,F} Q_F A_{1/2}(\tau_F) \right. \\ && \left. - \frac{ g_{h^0SS}
g^\prime_{ZSS} }{ m_S^2 } N_{c,S} Q_S A_0(\tau_S) \right ]
\end{eqnarray}
with $g^\prime_{ZXX}$ ($X=V,F,S$) standing for the coefficient of the $Z X^\ast X$ interaction. Note that in getting this expression,
we have neglected the $Z$ boson mass appeared in the loop functions since $m_{h^0}^2, m_X^2 \gg m_Z^2$, and consequently the
involved loop functions can be greatly simplified.

\item In the topflavor seesaw model, the decay $h^0 \to Z Z, W W^\ast$ are also induced by loop effects.  Their width expressions
are slightly complex, but can be still obtained in a way similar to that of
$h^0 \to \gamma \gamma$ if one neglects the vector boson mass appeared in the relevant loop functions.
Explicitly speaking, we have  \cite{ex-9}
\begin{eqnarray}
\Gamma_{h^0\to V V^\ast} &=&\delta_V \frac{G_F m_{h^0}^3}{16\pi \sqrt{2}}\frac{4 m_V^4}{m_{h^0}^4}
    \sqrt{\lambda(m_V^2,m_V^2;m_{h^0}^2)} \times \left[ A_V A_V^\ast \times \left( 2+\frac{(p_1 \cdot p_2)^2}{m_V^4}\right) \right. \nonumber \\
&& + (A_V B_V^\ast + A_V^\ast B_V) \times \left( \frac{(p_1 \cdot p_2)^3}{m_V^4} - p_1 \cdot p_2\right) \nonumber \\
&& \left. + B_V B_V^\ast \times \left( m_V^4 + \frac{(p_1 \cdot p_2)^4}{m_V^4} - 2(p_1 \cdot p_2)^2 \right) \right],
\end{eqnarray}
where $\delta_V=2(1)$ for $V=W(Z)$, $\lambda(x,y,z)=((z-x-y)^2-4xy)/z^2$ and
$p_1 \cdot p_2 = \frac{1}{2} \left( m_{h^0}^2 - 2 m_V^2 \right)$ with $m_V=m_W(m_Z)$ for $V=W(Z)$ respectively. The forms of $A_V$ and $B_V$ are
\begin{eqnarray}
\nonumber
A_V&=&
\frac{\alpha~ p_1 \cdot p_2 }{4 \pi m^2_V \delta_V} \frac{ m_W }{ g^3_2 \sin^2\theta_W} \times \left[ \frac{ g_{h^0 \tilde{V} \tilde{V}} g_{V \tilde{V} \tilde{V}^{\prime}}^2 }{ m^2_V } N_{c,V} A_{1}(\tau_{\tilde{V}}) \right . \nonumber \\
&& \left .  - \frac{ 2 g_{h^0 F F} g_{V F F^\prime}^2 }{ m_F } N_{c,F} A_{1/2}(\tau_F) - \frac{ g_{h^0 S S} g_{V S S^\prime}^2 }{m_S^2}  N_{c,S} A_0(\tau_S) \right ],
\nonumber \\
B_V&=& - \frac{ A_V }{ p_1 \cdot p_2 },
\end{eqnarray}
where the possible particles in the loops are $\tilde{V}, \tilde{V}^\prime = F_\mu^+, F_\mu^3$, $F, F^\prime = t_2$, $t_3$, $b_2$, $b_3$,
$\tau_2$, $\tau_3$, $\nu_{\tau_2}$, $\nu_{\tau_3}$ and $S, S^\prime = H^+, H^0$ respectively.
\end{itemize}

About above formulae, it should be noted that, if the $W$ and $Z$ mass appeared in the squared amplitudes and phase
spaces are also neglected, their expressions can be greatly simplified, and consequently they take similar forms.
In this case, their expressions in our model are approximated by
\begin{eqnarray}
\Gamma_{h^0 \to \gamma\gamma} &\simeq & \frac{\alpha^2 m_{h^0}^3}{1024 \pi^3}\Big|\frac{A_1(\tau_V)+ x A_0(\tau_S)}{v}+
  \frac{8\lambda_V  A_{\frac{1}{2}}(\tau_{F_1})}{3(\lambda_V v- m_H)}
   + \frac{8 \lambda_V A_{\frac{1}{2}}(\tau_{F_2})}{3(\lambda_V v+ m_H)} \Big|^2~, \nonumber \\
\Gamma_{h^0 \to gg} &\simeq & \frac{\alpha_s^2 m_{h^0}^3}{512 \pi^3}\Big|\frac{2\lambda_V A_{\frac{1}{2}}(\tau_{F_1})}{\lambda_V v- m_H}
  + \frac{2\lambda_V A_{\frac{1}{2}}(\tau_{F_2})}{\lambda_V v+ m_H} \Big|^2,\nn \\
\Gamma_{h^0 \to Z\gamma} & \simeq & \frac{\alpha^2 m_{h^0}^3}{512 \pi^3}\frac{1}{\sin^2 \theta_W}\Big|
\frac{\cos \theta_W }{v}A_1(\tau_V)+\frac{\cos \theta_W }{v} x A_0(\tau_S)\nonumber \\
& &~~~~~~~~~~~~~~~~~~~~ +\frac{1-\frac{4}{3} \sin^2 \theta_W}{\cos \theta_W}\[\frac{2\lambda_V A_{\frac{1}{2}}(\tau_{F_1})}{\lambda_V v- m_H}+
\frac{2\lambda_VA_{\frac{1}{2}}(\tau_{F_2})}{\lambda_V v+ m_H}\] \Big|^2, \nonumber \\
\Gamma_{h^0 \to ZZ} & \simeq & \frac{\alpha^2 m_{h^0}^3}{1024 \pi^3} \frac{1}{\sin^4 \theta_W} \Big|
\frac{\cos^2 \theta_W A_1(\tau_V)}{v}  + \frac{x \cos^2 \theta_W A_0(\tau_S)}{v}
\nonumber \\
& &
 ~~~~~~~~~~~~~~~~~~~+ \f{\lambda_V}{\cos^2 \theta_W}\sum\limits_{F}N_{ZZ}^F\[\frac{  A_{\frac{1}{2}}(\tau_{F_1})}
{ \lambda_V v-m_H} + \frac{ A_{\frac{1}{2}}(\tau_{F_2})}
{ \lambda_V v + m_H} \]\Big|^2, \nonumber \\
\Gamma_{h^0 \to W W^\ast} & \simeq & \frac{\alpha^2 m_{h^0}^3}{512 \pi^3} \frac{1}{\sin^4 \theta_W} \Big|
\frac{A_1(\tau_V)}{v} + \frac{x A_0(\tau_S)}{v} + \frac{2\lambda_V A_{\frac{1}{2}}(\tau_{F_1})}{\lambda_V v- m_H}
  + \frac{2\lambda_V A_{\frac{1}{2}}(\tau_{F_2})}{\lambda_V v+ m_H} \Big|^2
 \label{Simplified-expression}
\end{eqnarray}
where we have defined
\beqa
\tau_{F_1} &=& \f{4 (\lambda_V v - m_H)^2}{ m_{h^0}^2},~\tau_{F_2} =\f{ 4 (\lambda_V v + m_H)^2}{ m_{h^0}^2}~,
~\tau_V = \f{4 m_{F_{\mu}}^2}{ m_{h^0}^2}~,\tau_{S} =\f{ 4 m_{H^+}^2}{ m_{h^0}^2}~,\nn\\
\sum\limits_{F}N_{ZZ}^F &\equiv& \[3\big(\frac{1}{2}-\frac{2}{3}\sin^2 \theta_W \big)^2 + 3\big(-\frac{1}{2}+\frac{1}{3}\sin^2 \theta_W \big)^2
+ \big(-\frac{1}{2}+\sin^2 \theta_W \big)^2 + \frac{1}{4}\],  \nonumber
\eeqa
$\theta_W$ is the weak mixing angle and $x$ is introduced in Eq.(\ref{h-H+-H-}) to parameterize the $h^0 H^+ H^-$ and $h^0 H^0 H^0$ couplings.

From Eq.(\ref{Simplified-expression}), one can infer that if the decays are induced mainly by the vector-like fermions, $h^0 \to g g$
will be far dominant over the other decays. In this case, the diphoton rate in Eq.(\ref{cross-section}) is roughly determined by
$\Gamma_{h^0 \to \gamma \gamma}$, and Eq.(\ref{requirment}) is then equivalent to
\begin{eqnarray}
|I^{h^0}_{\gamma \gamma}| \simeq \frac{21.6 \pm 3.0}{{\rm TeV}}.  \label{diphoton-condition}
\end{eqnarray}
In the following, we will use this condition to find the solutions to the diphoton excess in the topflavor seesaw model.
Eq.(\ref{Simplified-expression}) also indicates that the branching ratios for the decays $h^0 \to W W^\ast, Z Z, Z \gamma$
are at least several times larger than that of $h \to \gamma \gamma$. As a result, the $h^0$ production can generate sizable
$WW^\ast$, $Z Z$ and $Z \gamma$ signals. Our model should be compatible with the LHC Run I constraints that are given by
\begin{eqnarray}
&& \sigma_{8{\rm TeV}}(pp\ra h^0 \ra Z\gamma)\leq 3.6 ~{\rm fb}, \quad \sigma_{8{\rm TeV}}(pp\ra h\ra Z Z)\leq 12 ~ {\rm fb},  \nonumber \\
&& \sigma_{8{\rm TeV}}(pp\ra h\ra WW^\ast)\leq 37~{\rm fb},  \quad \sigma_{8{\rm TeV}}(pp\ra h\ra g g)\leq 1.8~ {\rm pb}.
\end{eqnarray}

\subsection{Discussion and numerical results}

From Eq.(\ref{Simplified-expression}), one can learn that the involved parameters for the diphoton signal are
\begin{itemize}
\item the parameters in the scalar sector, which are $m_{h^0} = 750{\rm GeV}$, $v$, $x$ and $m_{H^+} = m_{H^0}$.
\item the parameter $\tan \theta$ in the gauge sector, which determines the flavoron mass.
\item the parameters in the fermion sector, which are $\lambda_V$ and $m_H$ used to determine
the fermion masses and their Yukawa couplings.
\end{itemize}
In order to illustrate our explanation of the excess in a concise way, we assume that all new particles other than $h^0$
are significantly heavier than $h^0$ so that $\tau_X=4m_X^2/m_{h^0}^2 \gg 1$. In this case, since
$A_1 (\tau_V)\simeq -7 $, $A_{\frac{1}{2}} (\tau_{F_i}) \simeq 4/3$ and $A_0 (\tau_S) \simeq 1/3$,  Eq.(\ref{diphoton-condition})
can be reexpressed by
\begin{eqnarray}
|\frac{-7}{v} +\frac{8}{3}( \frac{\lambda_V}{\lambda_V v - m_H} + \frac{\lambda_V}{\lambda_V v + m_H} ) \frac{4}{3}
+ \frac{x}{3 v} | \simeq \frac{20.7 \pm 2.8}{{\rm TeV}},  \label{approximation}
\end{eqnarray}
This equation reveals the following information
\begin{itemize}
\item The vector boson contribution interferes destructively with the fermion contribution. While for the scalar contribution,
it may interfere either constructively (if $x>0$) or destructively (if $x<0$) with the fermion contribution.
\item If $m_H \simeq 0$, the contribution from each fermion is usually significantly smaller than the vector boson contribution,
but the total fermion contribution in our model can cancel strongly with the vector contribution regardless the value of $\lambda_V$.
On the other hand, if $m_H$ is sufficiently large so that $\lambda_V v - m_H \ll \lambda_V v$ or equally speaking
$\lambda_V/(\lambda_V v - m_H) \gg 1/v$,
the fermion contribution may be dominant. This guides us to get the solution for the diphoton excess.
\item For $x \sim 1$, the scalar contribution is very small in comparison with the other contributions.
However, if $|x| \gg 1$, which is somewhat unnatural but still possible by tuning $\mu_1^2$ and $\mu_2^2$ to get $m_{H^+}$ in Eq.(\ref{m-H+}),
 the scalar contribution can be important.
\item For a large $v$, contributions from the vector boson and the scalar decrease quickly since they are proportional to $1/v^2$.
In contrast, if one keeps the lighter vector-like fermions at TeV scale by requiring $(\lambda_V v - m_H) \sim 1 {\rm TeV}$,
the vector fermion contribution can still be sizable even for a very large $v$.  In this case, the effective theory of our model at
TeV scale is similar to the minimal model mentioned in the Sec 1.
\item For $v = 10 {\rm TeV}$, $\lambda_V - m_H = 1 {\rm TeV}$ and $x=0$, we can make an estimation with Eq.(\ref{approximation}) that $\lambda_V \simeq 6.3 \pm 0.9$ can explain the diphoton excess at $1 \sigma$ level. The corresponding Yukawa coefficient for the $h^0 \bar{t_2} t_2$ interaction is $3 \pm 0.4$, which is about 3 times the top quark Yukawa coupling, but still significantly below the non-perturbative bound $4\pi/\sqrt{N_{c,F}}$.
\end{itemize}

\begin{figure}[t]
\centering
\includegraphics[width=7.5cm]{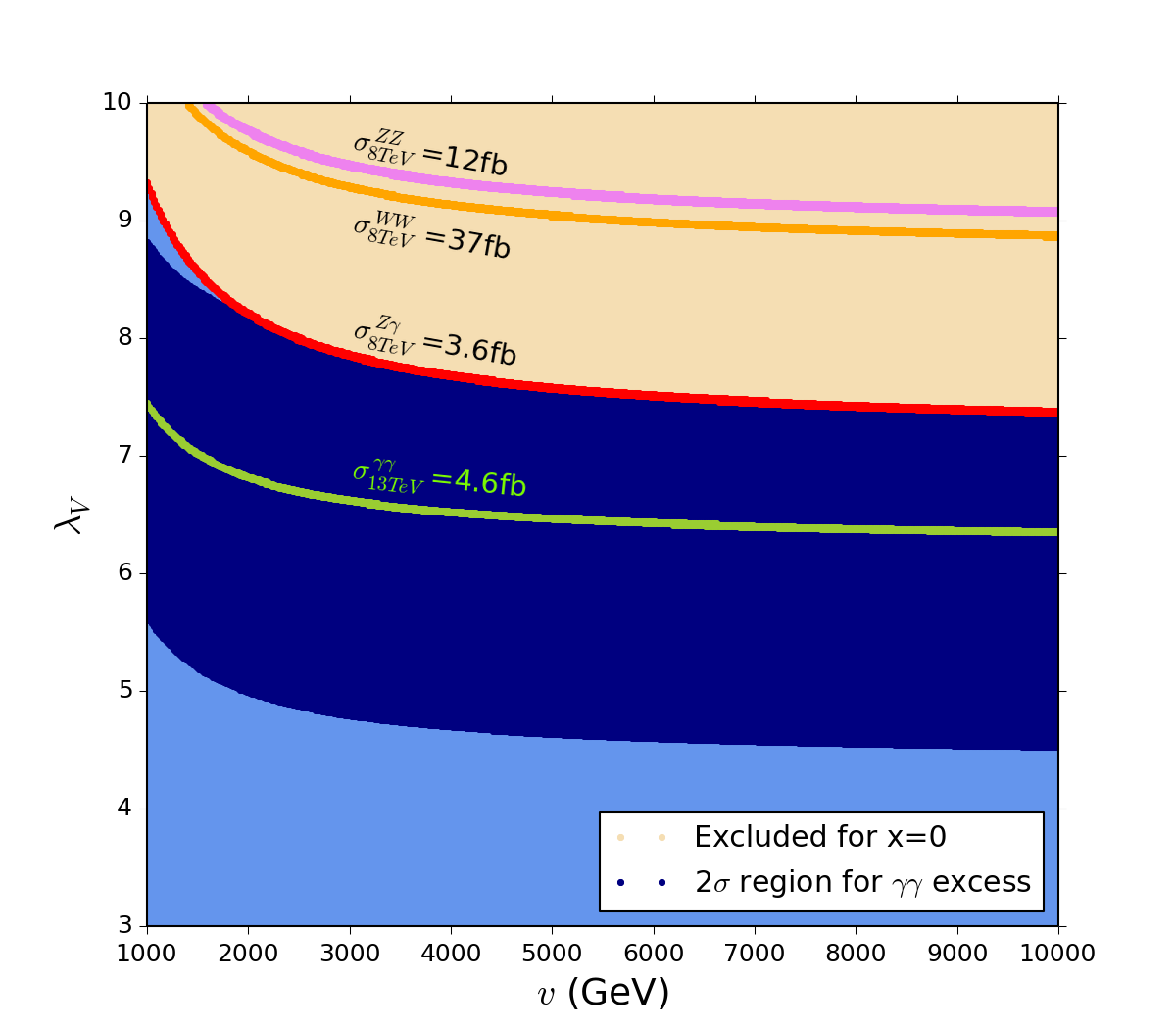} \includegraphics[width=7.5cm]{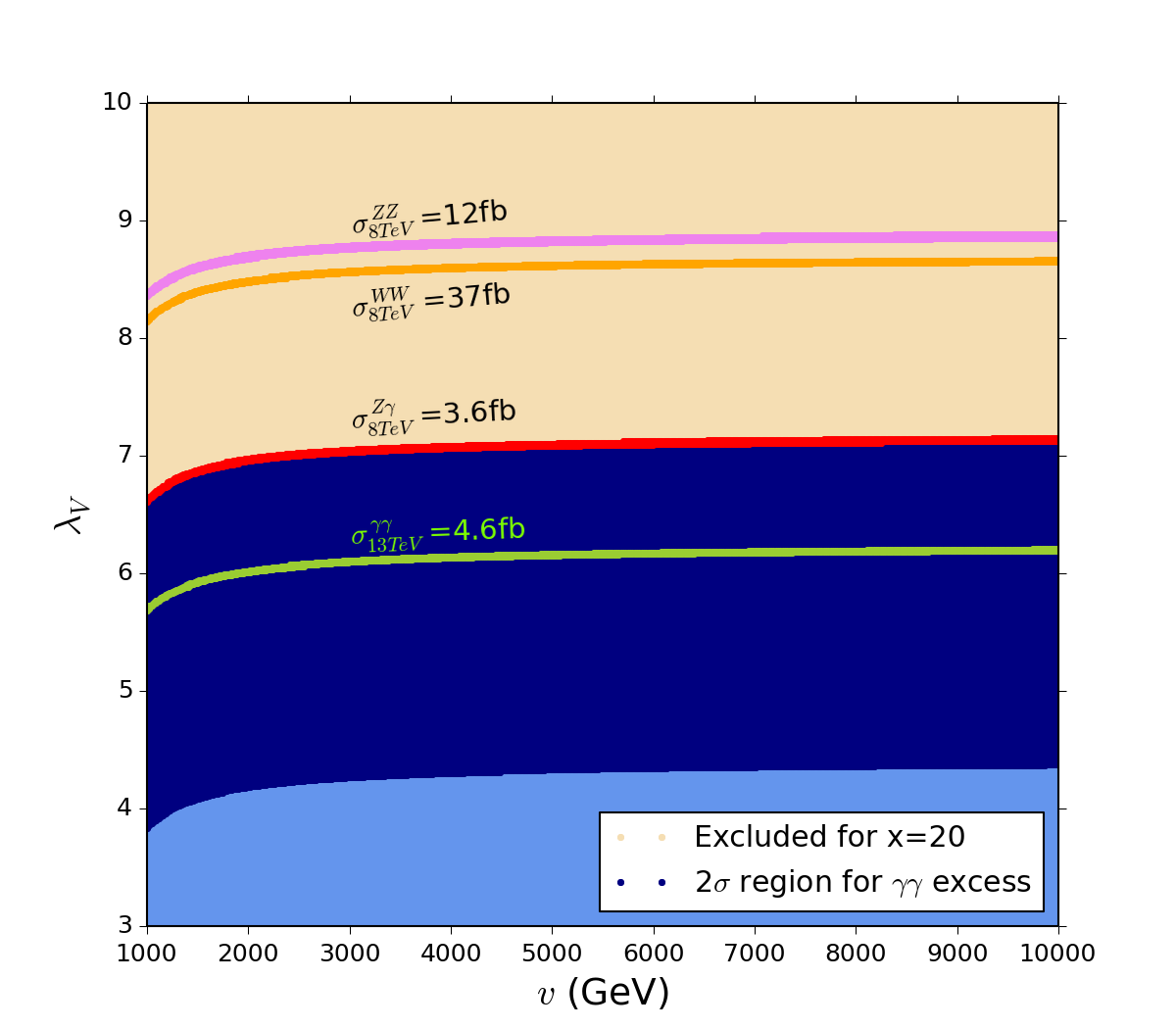}
\includegraphics[width=7.5cm]{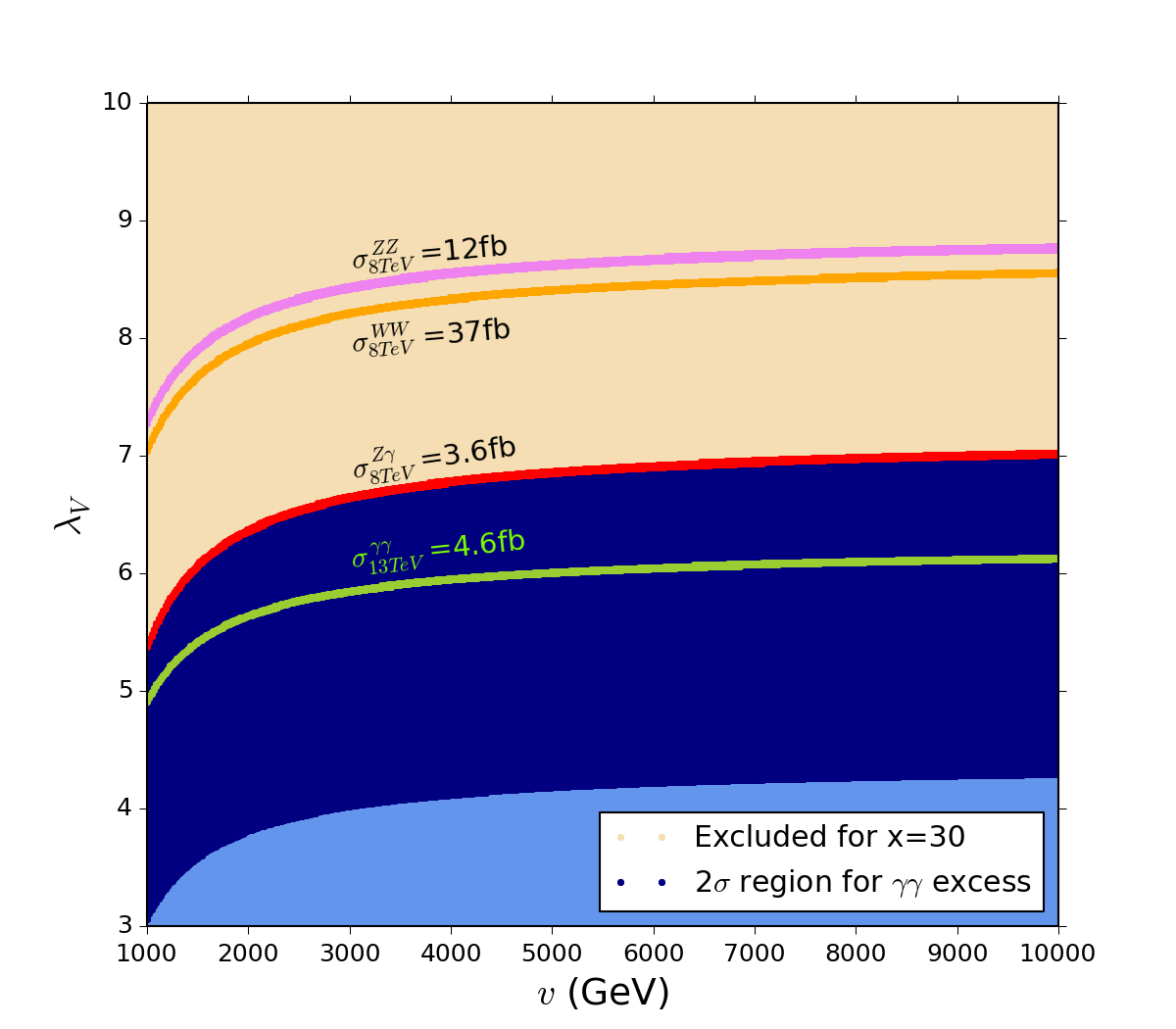} \includegraphics[width=7.5cm]{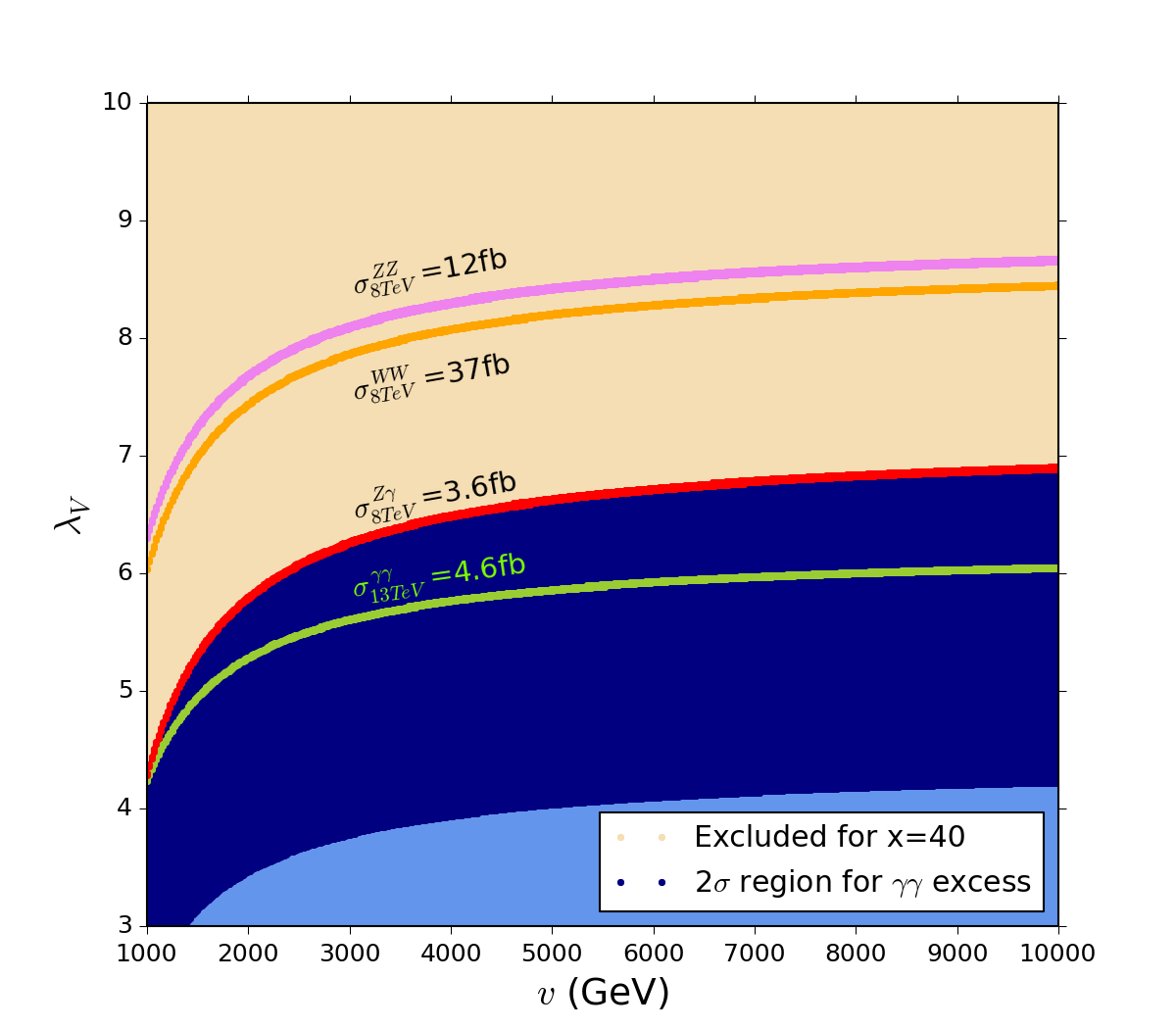}
\caption{Favored parameter space on the $\lambda_V-v$ planes for the diphoton excess with different choices of $x$, which parameterizes
the $h^0 H^+ H^-$ and $h^0 H^0 H^0$ coupling coefficient. The regions shaded by the blue color are able to explain the excess at $2 \sigma$ level,
and by contrast the regions covered by straw color are excluded by the upper bounds on $Z \gamma$ signal at the LHC Run I.
The blue lines correspond to $\sigma_{13 {\rm TeV}}^{\gamma \gamma} = 4.4~{\rm fb}$, which is the central value for the diphoton excess,
and the red lines, brown yellow lines and pink lines are the boundary lines of the $Z \gamma$,
$W W^\ast$ and $ZZ$ signals at the 8 TeV LHC respectively.   In getting this figure, we have fixed the masses of the
lighter vector-like fermions at $1 {\rm TeV}$, i.e. $m_{F_1} = \lambda_V v - m_H = 1 {\rm TeV}$, and $m_{H^+} = m_{H^0} = v$. In this case, the effective
theory of our model at $1 {\rm TeV}$ contains only vector-like fermions and $h^0$ for a sufficiently large $v$. Note that the Yukawa coupling coefficient for
the $h^0 \bar{F} F$ interaction is $\lambda_V/2$, instead of $\lambda_V$.  }
\label{fig1}
\end{figure}

In our numerical study, we fix $\tan \theta = 1$, $m_{h^0} = 750 {\rm GeV}$, $m_{H^+} = m_{H^0} = v$,
and $\lambda_V v - m_H = 1 {\rm TeV}$.
 We vary $\lambda_V$ and $v$ to get the favored parameter region for the diphoton excess  with $x=0, 20, 30, 40$ at each time.
The contours for $\sigma_{8 {\rm TeV}}^{Z \gamma} = 3.6~{\rm fb} $, $\sigma_{8 {\rm TeV}}^{Z Z} = 12~{\rm fb} $ and
$\sigma_{8 {\rm TeV}}^{W W^\ast} = 37~{\rm fb} $ in the $\la_v-v$ plane are also obtained. The corresponding results are shown in Fig.\ref{fig1},
where the upper panels are for the results with $x=0, 20$ and the lower panels correspond to the results
with $x=30,40$. From this figure, one can get following conclusions
\begin{itemize}
\item The topflavor seesaw model can explain the diphoton excess without conflicting with the constraints from the data
at the LHC Run I, and the central value of the excess can be obtained even for $v \sim 10 {\rm TeV}$.
\item Given a sufficiently large $v$, e.g. $v \gtrsim 6 {\rm TeV}$,  $\lambda_V \simeq 6$ is usually needed to predict
the central value of the excess. 
\item For $v \simeq 1 {\rm TeV}$ and $x = 40$, which corresponds to a tuning of $1/x$ in getting the mass of $H^+$,
$\lambda_V \simeq 4 $ is required to predict the central value of the excess.
 Moreover, to explain the excess at $2 \sigma$ level, the coupling can be as small as $\lambda_V \simeq 2.5$.
\item The LHC data at Run I has imposed rather tight constraints on our model. For the field configuration in our theory, the strongest
constraint comes from the upper bound on the $Z \gamma$ channel $\sigma_{8 {\rm TeV}}^{Z \gamma} \lesssim 3.6 {\rm fb}$, and it has
required $\sigma_{13 {\rm TeV}}^{\gamma \gamma} \lesssim 6 {\rm fb}$.  We remind that if we use
$\sigma_{8 {\rm TeV}}^{Z \gamma} \lesssim 6 {\rm fb}$ adopted in \cite{ex-0} as the constraint, the upper bound of
$\sigma_{13 {\rm TeV}}^{\gamma \gamma} $ becomes about $10 {\rm fb}$.
\end{itemize}

\section{Conclusion}
We propose to interpret the 750 GeV diphoton excess in a typical topflavor seesaw model. The new resonance $X$ can be identified as a CP-even scalar emerging from a certain bi-doublet higgs field. Such a scalar can couple to charged scalars, fermions as well as heavy gauge bosons predicted by the model, and as a result all of these particles contribute to the diphoton decay mode of the X.
Numerical analysis indicates that the model can predict the central value of the diphoton excess without contradicting
any constraints from 8 TeV LHC, and among the constraints,
the tightest one comes from the $Z \gamma$ channel, $\sigma_{8 {\rm TeV}}^{Z \gamma} \lesssim 3.6 {\rm fb}$, which requires
$\sigma_{13 {\rm TeV}}^{\gamma \gamma} \lesssim 6 {\rm fb}$ in most of the favored parameter space.

From theoretical point of view, our model has advantages in comparison with minimal frameworks. As we mentioned in Sec 1,
the key factor $\la_F/M_F$ for the diphoton rate in the minimal model scales like $1/v$ for a large $v$.
However, in our model the factor $\la_F/M_F$ is equal to
$c_v/v$ with $c_v\gg 1$. This is possible because the simply relation $M_F=\la_F v$ is unleashed and an
effective negative contribution to the vector-like fermions is generated to decrease $M_F$. Note that in the minimal model
imposing such a negative contribution by hand to spoil the relation $M_F=\la_F v$ is very unnatural.  Moreover, in minimal seesaw model with vector-like heavy
top quarks \cite{hjhe:topflavorseesaw}, the seesaw mechanism is fully responsible for the top quark mass. As a result,
there is undesirable strong correlation between the diphoton decay rate and the $h^0 \to \bar{t}t$ decay rate, and
the constraint from the $h^0$ mediated $\bar{t}t$ resonance at LHC Run I can falsify the model even though
the mixing between the top quark and heavy top ${\cal{T}}$ is very tiny.  In our model, however, the seesaw mechanism
contributes only a small portion of the top quark mass, and consequently there is no such a correlation.

\begin{acknowledgments}
 This work was supported by the Natural Science Foundation of China under grant numbers 11105124, 11222548, 90103013, 11275245 and 11575053; by the Open Project Program of State Key Laboratory of Theoretical Physics, Institute of Theoretical Physics, Chinese Academy of Sciences, P.R.
China (No.Y5KF121CJ1); by the Innovation Talent project of Henan Province under grant
number 15HASTIT017 and the Young-Talent Foundation of Zhengzhou University.
\end{acknowledgments}

\appendix

\section{The couplings needed in our calculation}
In the section, we enumerate the couplings needed in our calculation.

\subsection{The couplings involving the scalar $h^0$}

\begin{itemize}
\item The couplings of $h^0$ to gauge bosons.

These interactions comes from the kinetic term
 \beqa
 {\cal L}\supseteq Tr\[(D_\mu \Phi)^\da (D^\mu \Phi)\]~, \label{kinematic}
 \eeqa
and consequently, we have
\beqa
{\cal L}&\supseteq& \f{(h_1^2+h_2^2)v}{2}(F_\mu^3F^{\mu3}+2F_\mu^+F^{\mu-})h^0.  \label{h-F-F}
\eeqa

\item The couplings of $h^0$ to vector-like quarks.

These couplings are given by
\beqa
{\cal L}&\supseteq& - \frac{\la_V}{\sqrt{2}} \ \[ \rho_1 (\bar{{\cal{T}}}_L^1 {\cal{T}}^1_R + \bar{{\cal{T}}}_L^2 {\cal{T}}^2_R )   + \rho_3 (\bar{{\cal{B}}}_L^1 {\cal{B}}^1_R + \bar{{\cal{B}}}_L^2 {\cal{B}}^2_R )  \] + h.c.~,\nn\\
&\simeq & - \frac{\la_V}{2} \[ h^0 (\bar{t}_2 t_2 + \bar{t}_3 t_3 + \bar{b}_2 b_2 + \bar{b}_3 b_3 ) - H^0 (\bar{t}_2 t_2 + \bar{t}_3 t_3 - \bar{b}_2 b_2  - \bar{b}_3 b_3 ) \].
\eeqa
Note that there is an additional factor $\frac{1}{2}$ for the coupling coefficient. Also note that the vector-like leptons have same Yukawa
couplings as the quarks.

\item The couplings of $h^0$ to heavy scalars

These coupling originates from the $\Phi$ potential presented in Eq.(\ref{Bi-doublet-potential}). After tedious expansion of the $V(\Phi)$,
we find that they take following forms
\beqa
{\cal L}&\supseteq& - 2 (\lambda_1 - \lambda_2 - 2 \lambda_3 ) v h^0 (H^0 H^0 + 2 H^+H^-) \nonumber \\
&\equiv& - x \frac{m_{H^+}^2}{v} h^0 (\frac{1}{2} H^0 H^0 +  H^+H^-), \label{h-H+-H-}
\eeqa
where in the last step we introduce a dimensionless quantity $x$ to parameterize the interaction.
From Eq.(\ref{m-H+}), one can learn that $x=1$ if $2 \mu_2^2 = \mu_1^2$, and $x > 1$ ($x<1$) if
$2 \mu_2^2 < \mu_1^2$ ($2 \mu_2^2 > \mu_1^2$).

\end{itemize}

\subsection{The couplings of $W$ and $Z$ bosons to the heavy scalars}

These couplings originate from the kinematic term in Eq.(\ref{kinematic}), and the terms we will use are given by
\beqa
{\cal L}&\supset& -ig_2\[ (\pa^\mu H^-) W_\mu^+ H-(\pa^\mu H) W_\mu^+ H^- +(\pa^\mu H) W_\mu^- H^+-(\pa^\mu H^+) W_\mu^- H\.\nn\\
 &&~~~~+\left.(\pa^\mu H^-) W_\mu^3 H^+-(\pa^\mu H^+) W_\mu^3 H^-\] \nn\\
&&+\f{1}{2}g_2^2 g^{\mu \nu} \[2(W^3_\mu W^+_\nu H H^- + W^3_\mu W^-_\nu HH^+)-(W^+_\mu W^+_\nu H^-H^-+W^-_\mu W^-_\nu H^+H^+)\.\nn\\
&&~~~~~ +\left.2H^+H^-(W^3_\mu W^3_\nu + W^+_\mu W^-_\nu )+2W^+_\mu W^-_\nu H^2\].
\eeqa
The corresponding Feynman rules are
\bit
\item $H^-(p_1)-H^+(p_2)-Z^0_\mu(p_3): \quad \quad -i g_2\cos\theta_W(p_1-p_2)_\mu $,
\item $H^-(p_1)-H^+(p_2)-A_\mu(p_3): \quad \quad -i e (p_1-p_2)_\mu$,
\item $H(p_1)-H^+(p_2)-W_\mu^-(p_3): \quad \quad -i g_2(p_1-p_2)_\mu$,
\item $H(p_1)-H^-(p_2)-W_\mu^+(p_3): \quad \quad i g_2(p_1-p_2)_\mu$,
\item $ H^+-H^--W^+_\mu- W^-_\nu:  \quad \quad \quad \quad i g_2^2 g_{\mu \nu}$,
\item $ H-H-W^+_\mu- W^-_\nu:  \quad \quad \quad \quad \quad 2 i g_2^2 g_{\mu \nu}$,
\item $ H^+-H^--Z_\mu- Z_\nu:  \quad \quad \quad \quad \quad 2 i g_2^2 \cos^2 \theta_W g_{\mu \nu}$,
\item $ H^+-H^--Z_\mu- A_\nu:  \quad \quad \quad \quad \quad  2 i g_2^2  \sin\theta_W\cos\theta_W g_{\mu \nu}$,
\item $ H^+-H^--A_\mu- A_\nu:  \quad \quad \quad \quad  \quad 2 i e^2 g_{\mu \nu}$.
\eit
In getting the first four rules, we have defined the direction of the momentum as that pointing to the vertex.

\subsection{The couplings of $W$ and $Z$ bosons to the heavy fermions}

Denoting $F$ to be any of the fermion fields $t_2, t_3, b_2, b_3, \tau_2, \tau_3, \nu_{\tau_2}, \nu_{\tau_3}$, we have following Feynman rules
for $W$ and $Z$ bosons

\bit
\item $Z_\mu-F-F: \quad \quad \quad -i \f{g_2}{\cos\theta_W}\gamma_\mu(T_q^3- Q_q\sin^2\theta_W)$,
\item $W^+_\mu -t_i-b_j: \quad \quad \quad -i \f{\sqrt{2}}{2}g_2 \delta_{ij} \ga_\mu $,
\item $W^+_\mu -\tau_i-\nu_{\tau_j}: \quad \quad \quad -i \f{\sqrt{2}}{2}g_2 \delta_{ij} \ga_\mu $.
\eit

Moreover, we also find that the coupling of the $F^+ F^- Z$ interaction is same as that of the $W^+ W^- Z$ interaction in the SM,
and the coupling of the $F^+ W^- F^3$ interaction differs from that of the $W^+ W^- Z$ interaction by a factor of $1/\cos \theta_W$.


\begin{thebibliography}{99}
\vspace{-1mm}

\bibitem{ATLAS:750}  CMS note, CMS PAS EXO-15-004, "Search for new physics in high mass diphoton events in proton-proton collisions at 13 TeV".

\bibitem{CMS:750}   ATLAS note, ATLAS-CONF-2015-081, "Search for resonances decaying to photon pairs in 3.2 fb-1 of pp collisions at vs = 13 TeV with the ATLAS detector".

\bibitem{ex-0}
  D.~Buttazzo, A.~Greljo and D.~Marzocca,
  arXiv:1512.04929 [hep-ph].

\bibitem{ex-00}
  A.~Falkowski, O.~Slone and T.~Volansky,
  arXiv:1512.05777 [hep-ph].

\bibitem{ex-1}
K.~Harigaya and Y.~Nomura,
  arXiv:1512.04850 [hep-ph];
  Y.~Mambrini, G.~Arcadi and A.~Djouadi,
  arXiv:1512.04913 [hep-ph];
M.~Backovic, A.~Mariotti and D.~Redigolo,
  arXiv:1512.04917 [hep-ph];
 A.~Angelescu, A.~Djouadi and G.~Moreau,
  arXiv:1512.04921 [hep-ph];
  Y.~Nakai, R.~Sato and K.~Tobioka,
  arXiv:1512.04924 [hep-ph];
  S.~Knapen, T.~Melia, M.~Papucci and K.~Zurek,
  arXiv:1512.04928 [hep-ph];
 A.~Pilaftsis,
  arXiv:1512.04931 [hep-ph];
  R.~Franceschini {\it et al.},
  arXiv:1512.04933 [hep-ph];
  S.~Di Chiara, L.~Marzola and M.~Raidal,
  arXiv:1512.04939 [hep-ph].

\bibitem{ex-2}
  T.~Higaki, K.~S.~Jeong, N.~Kitajima and F.~Takahashi,
  arXiv:1512.05295 [hep-ph];
   S.~D.~McDermott, P.~Meade and H.~Ramani,
  arXiv:1512.05326 [hep-ph];
  J.~Ellis, {\it et al.},
  arXiv:1512.05327 [hep-ph];
  M.~Low, A.~Tesi and L.~T.~Wang,
  arXiv:1512.05328 [hep-ph];
  B.~Bellazzini, R.~Franceschini, F.~Sala and J.~Serra,
  arXiv:1512.05330 [hep-ph];
  R.~S.~Gupta, {\it et al.},
  arXiv:1512.05332 [hep-ph];
  C.~Petersson and R.~Torre,
  arXiv:1512.05333 [hep-ph];
  E.~Molinaro, F.~Sannino and N.~Vignaroli,
  arXiv:1512.05334 [hep-ph].

\bibitem{ex-3}
B.~Dutta,  {\it et al.},
  arXiv:1512.05439 [hep-ph];
  Q.~H.~Cao,  {\it et al.},
  arXiv:1512.05542 [hep-ph];
A.~Kobakhidze,  {\it et al.},
  arXiv:1512.05585 [hep-ph];
  S.~Matsuzaki and K.~Yamawaki,
  arXiv:1512.05564 [hep-ph];
  R.~Martinez, F.~Ochoa and C.~F.~Sierra,
  arXiv:1512.05617 [hep-ph];
  P.~Cox, A.~D.~Medina, T.~S.~Ray and A.~Spray,
  arXiv:1512.05618 [hep-ph];
  D.~Becirevic, E.~Bertuzzo, O.~Sumensari and R.~Z.~Funchal,
  arXiv:1512.05623 [hep-ph];
  J.~M.~No, V.~Sanz and J.~Setford,
  arXiv:1512.05700 [hep-ph];
  S.~V.~Demidov and D.~S.~Gorbunov,
  arXiv:1512.05723 [hep-ph];
  W.~Chao, R.~Huo and J.~H.~Yu,
  arXiv:1512.05738 [hep-ph];
  S.~Fichet, G.~von Gersdorff and C.~Royon,
  arXiv:1512.05751 [hep-ph];
  D.~Curtin and C.~B.~Verhaaren,
  arXiv:1512.05753 [hep-ph];
  L.~Bian, N.~Chen, D.~Liu and J.~Shu,
  arXiv:1512.05759 [hep-ph];
  J.~Chakrabortty,  {\it et al.},
  arXiv:1512.05767 [hep-ph];
  A.~Ahmed,  {\it et al.},
  arXiv:1512.05771 [hep-ph];
  P.~Agrawal,  {\it et al.},
  arXiv:1512.05775 [hep-ph];
  C.~Csaki, J.~Hubisz and J.~Terning,
  arXiv:1512.05776 [hep-ph];
  D.~Aloni,  {\it et al.},
  arXiv:1512.05778 [hep-ph];
  Y.~Bai, J.~Berger and R.~Lu,
  arXiv:1512.05779 [hep-ph].

\bibitem{ex-4}
  E.~Gabrielli, {\it et al.},
  arXiv:1512.05961 [hep-ph];
  R.~Benbrik, C.~H.~Chen and T.~Nomura,
  arXiv:1512.06028 [hep-ph];
  J.~S.~Kim, J.~Reuter, K.~Rolbiecki and R.~R.~de Austri,
  arXiv:1512.06083 [hep-ph];
  A.~Alves, A.~G.~Dias and K.~Sinha,
  arXiv:1512.06091 [hep-ph];
 E.~Megias, O.~Pujolas and M.~Quiros,
  arXiv:1512.06106 [hep-ph];
 L.~M.~Carpenter, R.~Colburn and J.~Goodman,
  arXiv:1512.06107 [hep-ph];
 J.~Bernon and C.~Smith,
  arXiv:1512.06113 [hep-ph];
\bibitem{ex-5}
  D.~Barducci, A.~Goudelis, S.~Kulkarni and D.~Sengupta,
  arXiv:1512.06842 [hep-ph];
  M.~Chala, M.~Duerr, F.~Kahlhoefer and K.~Schmidt-Hoberg,
  arXiv:1512.06833 [hep-ph];
    M.~Bauer and M.~Neubert,
  arXiv:1512.06828 [hep-ph];
  J.~M.~Cline and Z.~Liu,
  arXiv:1512.06827 [hep-ph];
    W.~S.~Cho, D.~Kim, K.~Kong, S.~H.~Lim, K.~T.~Matchev, J.~C.~Park and M.~Park,
  arXiv:1512.06824 [hep-ph];
    L.~Berthier, J.~M.~Cline, W.~Shepherd and M.~Trott,
  arXiv:1512.06799 [hep-ph];
    J.~S.~Kim, K.~Rolbiecki and R.~R.~de Austri,
  arXiv:1512.06797 [hep-ph];
    X.~J.~Bi, Q.~F.~Xiang, P.~F.~Yin and Z.~H.~Yu,
  arXiv:1512.06787 [hep-ph];
    M.~Dhuria and G.~Goswami,
  arXiv:1512.06782 [hep-ph];
    J.~J.~Heckman,
  arXiv:1512.06773 [hep-ph];
    W.~Liao and H.~q.~Zheng,
  arXiv:1512.06741 [hep-ph];
    F.~P.~Huang, C.~S.~Li, Z.~L.~Liu and Y.~Wang,
  arXiv:1512.06732 [hep-ph];
  F.~Wang, L.~Wu, J.~M.~Yang and M.~Zhang,
  arXiv:1512.06715 [hep-ph];
    T.~F.~Feng, X.~Q.~Li, H.~B.~Zhang and S.~M.~Zhao,
  arXiv:1512.06696 [hep-ph];
    D.~Bardhan, D.~Bhatia, A.~Chakraborty, U.~Maitra, S.~Raychaudhuri and T.~Samui,
  arXiv:1512.06674 [hep-ph];
    J.~Chang, K.~Cheung and C.~T.~Lu,
  arXiv:1512.06671 [hep-ph];
  M.~x.~Luo, K.~Wang, T.~Xu, L.~Zhang and G.~Zhu,
  arXiv:1512.06670 [hep-ph];
    X.~F.~Han and L.~Wang,
  arXiv:1512.06587 [hep-ph];
    H.~Han, S.~Wang and S.~Zheng,
  arXiv:1512.06562 [hep-ph];
    R.~Ding, L.~Huang, T.~Li and B.~Zhu,
  arXiv:1512.06560 [hep-ph];
    I.~Chakraborty and A.~Kundu,
  arXiv:1512.06508 [hep-ph];
    C.~Han, H.~M.~Lee, M.~Park and V.~Sanz,
  arXiv:1512.06376 [hep-ph];
    M.~T.~Arun and P.~Saha,
  arXiv:1512.06335 [hep-ph];
    W.~Chao,
  arXiv:1512.06297 [hep-ph].


\bibitem{ex-6}
  P.~S.~B.~Dev and D.~Teresi,
  arXiv:1512.07243 [hep-ph];
    A.~Belyaev, G.~Cacciapaglia, H.~Cai, T.~Flacke, A.~Parolini and H.~Serodio,
  arXiv:1512.07242 [hep-ph];
    J.~de Blas, J.~Santiago and R.~Vega-Morales,
  arXiv:1512.07229 [hep-ph];
    G.~M.~Pelaggi, A.~Strumia and E.~Vigiani,
  arXiv:1512.07225 [hep-ph];
    U.~K.~Dey, S.~Mohanty and G.~Tomar,
  arXiv:1512.07212 [hep-ph];
    A.~E.~C.~Hernandez and I.~Nisandzic,
  arXiv:1512.07165 [hep-ph];
    C.~W.~Murphy,
  arXiv:1512.06976 [hep-ph];
    S.~M.~Boucenna, S.~Morisi and A.~Vicente,
  arXiv:1512.06878 [hep-ph].

\bibitem{ex-7}
  J.~Gu and Z.~Liu,
  arXiv:1512.07624 [hep-ph];
    M.~Cvetic, J.~Halverson and P.~Langacker,
  arXiv:1512.07622 [hep-ph];
    W.~Altmannshofer, J.~Galloway, S.~Gori, A.~L.~Kagan, A.~Martin and J.~Zupan,
  arXiv:1512.07616 [hep-ph];
    Q.~H.~Cao, S.~L.~Chen and P.~H.~Gu,
  arXiv:1512.07541 [hep-ph];
    S.~Chakraborty, A.~Chakraborty and S.~Raychaudhuri,
  arXiv:1512.07527 [hep-ph];
    M.~Badziak,
  arXiv:1512.07497 [hep-ph];
    K.~M.~Patel and P.~Sharma,
  arXiv:1512.07468 [hep-ph];
    S.~Moretti and K.~Yagyu,
  arXiv:1512.07462 [hep-ph];
    W.~C.~Huang, Y.~L.~S.~Tsai and T.~C.~Yuan,
  arXiv:1512.07268 [hep-ph].

\bibitem{ex-8}
  L.~J.~Hall, K.~Harigaya and Y.~Nomura,
  arXiv:1512.07904 [hep-ph];
    J.~A.~Casas, J.~R.~Espinosa and J.~M.~Moreno,
  arXiv:1512.07895 [hep-ph];
    J.~Zhang and S.~Zhou,
  arXiv:1512.07889 [hep-ph];
    J.~Liu, X.~P.~Wang and W.~Xue,
  arXiv:1512.07885 [hep-ph];
    K.~Cheung, P.~Ko, J.~S.~Lee, J.~Park and P.~Y.~Tseng,
  arXiv:1512.07853 [hep-ph];
    K.~Das and S.~K.~Rai,
  arXiv:1512.07789 [hep-ph];
    H.~Davoudiasl and C.~Zhang,
  arXiv:1512.07672 [hep-ph];
    B.~C.~Allanach, P.~S.~B.~Dev, S.~A.~Renner and K.~Sakurai,
  arXiv:1512.07645 [hep-ph];
    N.~Craig, P.~Draper, C.~Kilic and S.~Thomas,
  arXiv:1512.07733 [hep-ph];
  Y.~Hamada, T.~Noumi, S.~Sun and G.~Shiu,
  arXiv:1512.08984 [hep-ph].

\bibitem{ex-9}
   J.~Cao, C.~Han, L.~Shang, W.~Su, J.~M.~Yang and Y.~Zhang,
  arXiv:1512.06728 [hep-ph].

\bibitem{vacuum}
  J.~Zhang and S.~Zhou,
  arXiv:1512.07889 [hep-ph];
  M.~Son and A.~Urbano,
  arXiv:1512.08307 [hep-ph];
  J.~Cao, L.~Shang, W.~Su, Y.~Zhang and J.~Zhu,
  arXiv:1601.02570 [hep-ph].

\bibitem{tcond} V. Miransky, M. Tanabashi, and K. Yamawaki, Phys.Lett. B221, 177 (1989);\\
   V. Miransky, M. Tanabashi, and K. Yamawaki, Mod.Phys.Lett. A4, 1043 (1989);\\
   W. Marciano, Phys.Rev.Lett. 62, 2793 (1989);\\
   W. J. Marciano, Phys.Rev. D41, 219 (1990);\\
   W. A. Bardeen, C. T. Hill, and M. Lindner, Phys.Rev. D41, 1647 (1990).

\bibitem{topcolor}  C.T. Hill, hep-ph/9702320, hep-ph/9802216;
                    G. Cvetic, Rev. Mod. Phys. 71, 513 (1999).

\bibitem{topseesaw} B. A. Dobrescu and C. T. Hill, Phys. Rev. Lett. 81, 2634 (1998);\\
                 R. S. Chivukula, B. A. Dobrescu, H. Georgi and C. T. Hill, Phys. Rev. D 59, 075003 (1999);\\
                 B. A. Dobrescu, Phys. Rev. D 63, 015004 (2001).

\bibitem{HHT} H.-J. He, C. T. Hill and T. M. P. Tait, Phys. Rev. D 65, 055006 (2002).

\bibitem{topflavor} E. Malkawi, T. Tait, C.-P. Yuan, Phys. Lett. B385, 304 (1996);
                   D. Muller and S. Nandi, ibid. B383, 345 (1996).

\bibitem{flavor-dynamics}
  E.~Malkawi, T.~M.~P.~Tait and C.~P.~Yuan,
  Phys.\ Lett.\ B {\bf 385}, 304 (1996)
  doi:10.1016/0370-2693(96)00859-3
  [hep-ph/9603349].

\bibitem{hjhe:topflavorseesaw} H.-J. He, T.M.P. Tait, C.-P. Yuan, Phys.\ Rev.\ D62:011702 (2000).

\bibitem{Bi-doublet}
  N.~G.~Deshpande, J.~F.~Gunion, B.~Kayser and F.~I.~Olness,
  Phys.\ Rev.\ D {\bf 44}, 837 (1991).
  doi:10.1103/PhysRevD.44.837


\bibitem{ATLAS:125} G. Aad et al.(ATLAS Collaboration), Phys. Lett. B710, 49 (2012).

\bibitem{CMS:125} S. Chatrachyan et al.(CMS Collaboration), Phys. Lett.B710, 26 (2012).

\bibitem{vector-like-quark-search}
  [ATLAS Collaboration],
  Phys.\ Rev.\ D {\bf 91} (2015) 11,  112011
  [arXiv:1503.05425 [hep-ex]];
  [ATLAS Collaboration],
  JHEP {\bf 1508} (2015) 105
  [arXiv:1505.04306 [hep-ex]];
  CMS Collaboration [CMS Collaboration],
  CMS-PAS-B2G-15-006.

\bibitem{giudice} Roberto Franceschini, {\it et. al.}  arXiv: 1512.04933 [hep-ph].

\bibitem{HCS} https://twiki.cern.ch/twiki/bin/view/LHCPhysics/CERNYellowReportPageAt1314TeV

\bibitem{carena} Marcela Carena, Ian Low, Carlos E. M. Wagner, JHEP08(2012)060.

\end{thebibliography}
\end{document}